\documentclass{siam-wns-article}
\usepackage{times}
\usepackage{indentfirst}
\usepackage{graphicx}
\usepackage{amsthm}

\title{Computing Shortest Paths using A*, Landmarks, and Polygon Inequalities}

\author{Newton H. Campbell Jr., Raytheon BBN Technologies}

\usehyperref 

\begin{document}
\maketitle

\section*{Summary}
We introduce a new dual-landmark heuristic for the A* algorithm that references a data structure of size $\theta(|\textbf{\emph{L}}|^{2} + |V|)$, where $\textbf{\emph{L}}$ represents a set of strategically chosen landmark vertices and $V$ the set of vertices in the graph. This heuristic's benefits are permitted by a new approach for computing lower bounds based on generalized polygon inequalities, in which each landmark stores the distances between it and vertices within its graph partition. In this paper, we demonstrate experimental results that show exemplify the space complexity reduction and speedup in comparison to its single-landmark counterpart for large graphs.
\section{Introduction}
Modern navigation planning requires the ability to regularly compute the shortest path between two points in massive road networks. In such cases, preprocessing algorithms are used to increase the performance of shortest path  queries. Many such algorithms require heavy upfront computation and storage. Few algorithms concern themselves with the space complexity required to aid queries. The problem that this research addresses is that modern shortest path preprocessing algorithms have space and preprocessing time requirements for large-scale graphs that are impractical for real-world applications.
\section{A*, Landmarks, and Polygon Inequalities}
ALT \cite{GoldbergHarrelson2005} describes a preprocessing technique for shortest path queries that, prior to query time, chooses a relatively small number of landmark nodes in a graph and computes the distances between all vertices and these landmarks, allowing the A* algorithm to leverage the triangle inequality during search queries. The algorithm works as follows:
For a simple graph $G$ with vertices $A,l,C \in V$, where $l$ is a landmark vertex chosen beforehand, the shortest path distances between each vertex allow the graph to form a metric space. Therefore, for the distances between vertices $A,l,C \in V$, the following reverse triangle inequality holds:
\begin{equation}
    \label{triangle_inequality_lower_bound_abl}
    |d(A,l)-d(l,C)| \le d(A,C)
\end{equation}
ALT uses this inequality to create a heuristic for A* that estimates the distance between vertices $A$ and $C$. By computing and storing the values between each chosen landmark and all vertices in the graph prior to performing any shortest path queries, this estimate is performed for each computed landmark upon a visit to vertex $A$. The maximum of these estimates is the heuristic function's value, denoted as $\pi_t^{L}$.
\par
\begin{figure*}
\centering
\includegraphics[scale=0.7]{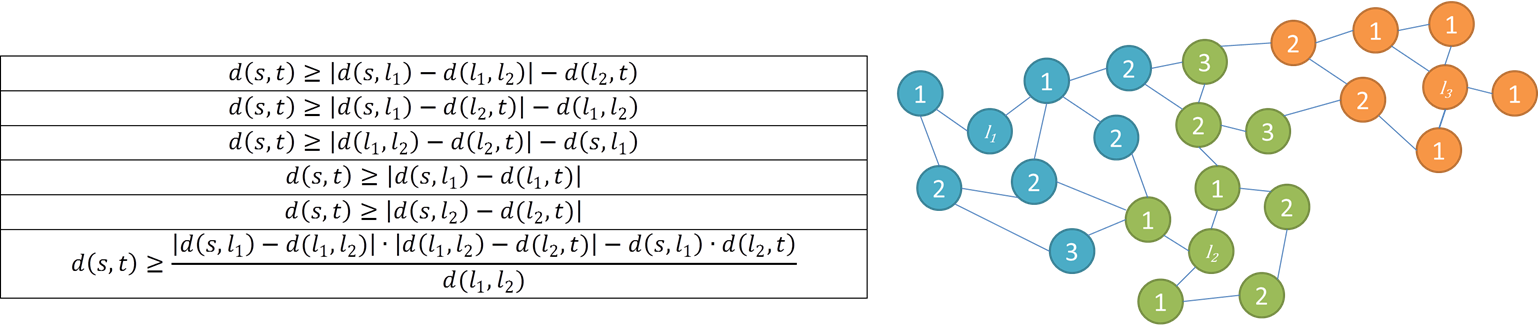}
\caption{Dual landmark ALP heuristic inequalities(left) and a labeling example for distributed embedding on a graph (right)}
\label{fig:equations_example}
\end{figure*}
\par
By using information about multiple landmarks, new lower bounds can be computed from either generalized polygon inequalities or ones specific to any shape embedded within the graph. The use of these new lower bounds as a heuristic has resulted in a new class of algorithms called ALP, for A*, Landmarks, and Polygon Inequalities. The base case for this class of algorithms is the heuristic used for the ALT algorithm. Here, we demonstrate that polygon inequalities for quadrilaterals can also be used to establish the lower bounds for shortest path queries in a graph. The reverse quadrilateral inequalities shown in Figure \ref{fig:equations_example} hold for a graph $G$ with source and target vertices $s,t \in V$ and chosen landmarks $l_1,l_2 \in V$. The first five are derived from the triangle inequality as applied to quadrilaterals. A potential problem with these inequalities is that they have ability to generate negative lower bound estimates. However, because multiple points are used, a varying set of inequalities can be generated to estimate distances. When attempting to estimate lower bounds using ALP, other inequalities should be considered such that the highest possible estimate can be used. We use the sixth equation, derived from Ptolemy's inequality, as a concrete example for quadrilaterals.
\par
Just as with ALT, the maximum over the set of these lower bounds are used to tighten the lower bound for the distance between two vertices. We denote the maximum of the six equations for ALP as $\pi_t^{DL}$, ALP's heuristic for A*. The following theorems then apply to $\pi_t^{DL}$:
\begin{description}
  \item Theorem 1: $\pi_t^{DL}$ is an admissible heuristic.
  \item Theorem 2: Using distributed embedding, $\pi_t^{DL}$ is not consistent.
  \item Theorem 3: $\pi_t^{DL}$ does not dominate $\pi_t^{L}$ over the same set of landmarks.
  \item Theorem 4: $\pi_t^{L}$ does not dominate $\pi_t^{DL}$ over a different set of landmarks.
\end{description}
\subsection{Distributed Embedding}
ALP's data structure can exhibit a space complexity of $\theta(|\textbf{\emph{L}}|^2 + |V|)$ (as opposed to ALT's $\theta(|\textbf{\emph{L}}| \cdot |V|)$)using the following technique, called \emph{distributed embedding}. With a partitioned graph as input, the dual landmark approach identifies a single landmark within each partition and computes a shortest path tree for the subgraph induced by each chosen landmark's graph partition. Each vertex in the graph is labeled with an identifier, signifying its landmark partition and the distance to its corresponding landmark. Any of the landmark selection methods for ALT can be used for the subgraph induced by the graph partition to select an optimal set. The final step of this process is an all-pairs shortest path calculation among the landmark set. An example of distributed embedding is illustrated in Figure \ref{fig:equations_example}. Allowing each landmark in the graph to access only a subset of the graph limits the size of the data structure used at query time as well as bounding the number of operations performed to actually compute the heuristic.
\section{Results}
\begin{figure}
\centering
\includegraphics[scale=0.30]{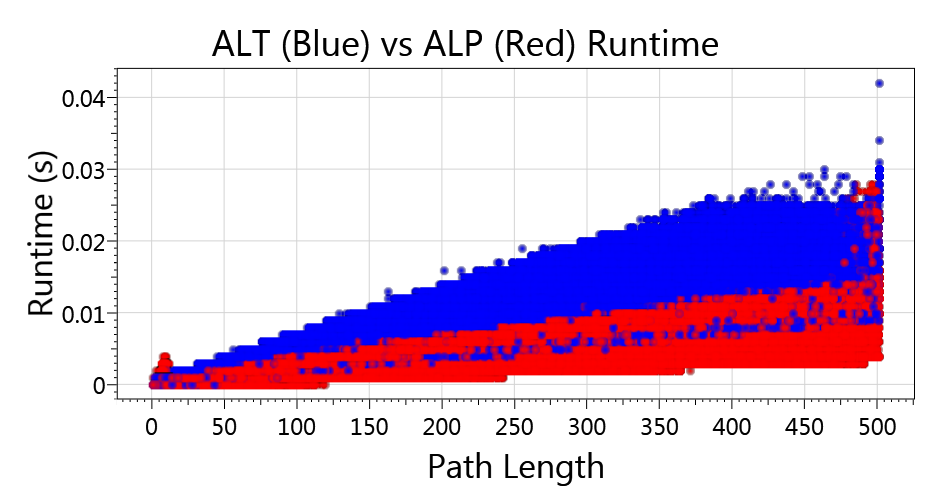}
\caption{Graph demonstrating ALT (Blue) performance in contrast to ALP (Red) performance for growing path lengths}
\label{fig:ALTvALP}
\end{figure}

Experimentation was performed as an initial investigation of the ALP dual landmark heuristic's behavior and its performance bounds in comparison to ALT. Comparison between ALT and ALP was performed using experimental benchmark road data from DIMACS and all available large-scale graphs (of size $10^6$ nodes) that can be generated by NetworkX 1.9. Random landmark selection was used for each trial run of the two algorithms on these datasets. The Louvain algorithm \cite{ICT4DBibliography2429} was used for the partitioning of each graph prior to distributed embedding. As illustrated in Figure \ref{fig:ALTvALP}, queries for paths with distances between 1 and 501 were called $10^5$ times.
\subsection{Discussion}
We have studied the effects of using an ALP data structure of size $\theta(|\textbf{\emph{L}}|^{2} + |V|)$ as opposed to ALT's previous $\theta(|\textbf{\emph{L}}|\cdot|V|)$. The resulting data show significant improvement of the runtime of the dual landmark ALP heuristic over the ALT heuristic on a diverse set of graphs with larger path lengths, as well as a significant reduction in required memory. In general, such geometric heuristics can be derived by identifying any polygon in a graph and setting the heuristic values for A* equal to the maximum derived lower bound of one of its sides. Future research will highlight the benefits and drawbacks to different landmark selection techniques for ALP and, if possible, leverage the distributed embedding method to create new forms of landmark selection that are possible only in this environment.
\section{Acknowledgments}
Special thanks to advisor Dr. Michael J. Laszlo, my doctoral committee at the Nova Southeastern University GSCIS and my employer, Raytheon BBN Technologies for their support throughout the Computer Science PhD program.

\bibliographystyle{abbrv}
\bibliography{NCampbellSIAM2015}

\end{document}